**Interface identification of the solid electrolyte interphase on graphite.**

Elena Zvereva[*,a,b,c], Damien Caliste[a,b] and Pascal Pochet[*,a,b].

[a] Université Grenoble Alpes, CS 40700, 38058 Grenoble, France.

[b] Atomistic Simulation Lab (L_Sim), MEM, INAC, CEA, rue des Martyrs 17, 38054 Grenoble, France.

[c] A.E. Arbuzov Institute of Organic and Physical Chemistry, Kazan Scientific Centre, Russian Academy of Sciences, Arbuzov street 8, 420088 Kazan, Russia.

**Abstract.**

By means of Density Functional Theory calculations we evaluate several lithium carbonate - graphite interface models as a prototype of the Solid Electrolyte Interphase capping layer on graphite anodes in lithium-ion batteries. It is found that only an ($a,b$)-oriented $Li_2CO_3$ slab promotes tight binding with graphite. Such mutual organization of the components combines their structural features and reproduces coordination environment of ions, resulting in an adhesive energy of 116 meV/Å$^2$ between graphite and lithium carbonate. This model also presents a high potential affinity with bulk. The corresponding charge distribution at such interface induces an electric potential gradient, such a gradient having been experimentally observed. We regard the mentioned criteria as the key descriptors of the interface stability and recommend them as the principal assessments for such interface study. In addition, we evaluate the impact of lithiated graphite on the stability of the model interface and study the generation of different point defects as mediators for Li interface transport. It is found that Li diffusion is mainly provided by interstitials. The induced potential gradient fundamentally assists the intercalation up to lithiation ratio of 70%.

---
[*] Corresponding authors, tel.: +334 38 78 63 65,
E-mail addresses: zverevaelena@ymail.com (Elena Zvereva), pascal.pochet@cea.fr (Pascal Pochet).

**1. Introduction.**

Lithium-ion batteries (LIBs) play an important role in rechargeable battery technologies due to such advantages as high energy storage, low self-discharge, reduced gas emission, small memory effect. In these batteries, the electrolyte reduction on graphite anode leads to the formation of a solid electrolyte interphase (SEI) that adheres well to the anode and prevents further electrolyte decomposition while simultaneously participating in Li-ion transport from electrolyte to anode [1-5]. Over the past decades various experimental techniques have revealed that the molecular composition of the SEI depends on the type of electrolyte, anode material and lithium salts [6,7]. It consists of different components such as $Li_2CO_3$, $Li_2O$, LiF, organic and polymeric species [8]. Besides, it is known that the SEI comprises a dense layer of inorganic components close to the graphite anode, followed by a porous organic or polymeric layer in the vicinity of the electrolyte phase [9]. The thickness of the SEI varies from a few angstroms to hundreds of nanometers [8, 10-11] and is influenced by the formation temperature [12-13] and applied voltage scan rate [14]. In addition to experimental methods, *ab initio* and Molecular Dynamics simulations have been used to study electrolyte decomposition [15-16], structural organization of the molecules at the electrode surface [17], and to elucidate the initial stages of SEI formation [16, 18-21] and Li diffusion mechanisms [21-25].

None of the existing studies, both experimental or numerical, gives a clear insight on what is the structural composition of an already formed SEI and the steric orientation of its components near the graphite anode. The nature of the SEI - graphite interface and the interactions which enable its adhesion are unknown. Moreover, a model interface cannot be considered just as the relaxation of space-distributed compounds. Indeed, the complex reorganizations at the interface, due to chemical and physical interactions, are difficult to capture by molecular dynamic simulations because of the limited time-scale that can be probed. All these difficulties cause that no reasonable atomistic model of the interface between the SEI and graphite has been proposed yet.

In this paper we present a stable SEI – graphite interface model at the atomic scale. The SEI is represented by $Li_2CO_3$ as the latter is one of the main stable components when ethylene carbonate-based electrolytes are used [5, 12, 19]. Within Density Functional Theory (DFT) framework, we study different possibilities of $Li_2CO_3$ - graphite arrangement with the aim to clarify organization and adhesion of the SEI on the graphite anode and elucidate their mutual impact on Li transport. We demonstrate that the stability of the model interface strongly depends on good connection between the two materials based on successful combination of their structural and electronic features. Moreover, the same features drive the Li migration toward the graphite galleries.

**2. Computational Details.**

All calculations were performed with the BigDFT code based on a real-space wavelet basis set [26]. The Perdew-Burke-Ernzerhof (PBE) [27] density functional was used for approximating the exchange correlation potential. Core electrons were treated with the Hartwigsen-Goedecker-Hutter [28] pseudo potentials with the Krack variant [29]. The convergence within the wavelet basis-set was checked with respect to the grid spacing, the chosen value of 0.42 bohr providing an accuracy of 1 meV/atom. Total energy and its derivatives were obtained using a direct minimization scheme bringing the gradient lower than $10^{-4}$ atomic units. Finally, atomic relaxations were carried out with the FIRE algorithm [30] until the forces acting on each atom were less than 0.02 eV/Å. For a test purpose a stricter convergence was performed, bringing forces lower than $5.10^{-4}$ eV/Å, for the main model of interface presented thereafter. The coordination values reported in Table 1 are not changed by this additional relaxation. We thus consider that our analysis of SEI interface with graphite is based on relaxed enough structures.

All considered systems are electrically neutral, periodic in two directions and free in the third one (Surface Boundary Conditions, SBC), *i.e.* there is no need for vacuum region above the surfaces unlike with the Periodic Boundary Conditions. Computing in SBC super-cells is possible thanks to the real-space formalism of the basis-set and an exact treatment of the 2D periodicity in the Poisson solver [31]. In addition, this SBC solver allows the direct treatment of dipolar interface which is crucial in this study. All simulated super-cells contain around 400 atoms, and thus no k-point mesh was used. These specific sizes were chosen to accommodate the monoclinic $Li_2CO_3$ phase on top of graphite with minimal shearing of the lithium carbonate phase while keeping tractable DFT calculations (see section "Modification of a lithium carbonate unit cell" of Supporting Information).

In non-covalently bound systems like graphite or $Li_2CO_3$, the correct treatment of nonlocal London dispersion interactions is mandatory for accurate geometries and binding energies. These interactions are not included in any Generalized Gradient Approximation density functionals and require dispersion corrections. Here we used the DFT-D2 London dispersion correction scheme proposed by Stefan Grimme [32]. This approach implemented in BigDFT and combined with the wavelet basis set is shown to provide results close to those of CCSD(T) benchmark quality [33].

To compare the respective stability of various model interfaces, with different compositions, we define surface formation energy ($E_S$) as the energetic cost per surface unit to build the interface from individual components:

$$E_S = (E_{Total} - \sum_i \alpha_i \mu_i) / A_{interface} \qquad (1)$$

where $E_{Total}$ is the total energy of the system, $\mu_i$ are the chemical potentials of the used chemical species, $\alpha_i$ are the coefficients required to equilibrate the composition between the interface system and the reference components, and $A_{interface}$ is the planar surface area cutting the super-cell at the interface. The main reference components are bulk graphite and bulk lithium carbonate. In some cases metallic lithium is also used to handle off-stoichiometry composition. This particular choice of chemical potentials allow to easily compare the relative stability of different models with varying stoichiometries but it does not provide any information on the adhesion of lithium carbonate on top of graphite. For this concern, we have used a simpler formula:

$$E_{Adh} = (E_{Total} - E_{Slab\_LithiumCabonate} - E_{Slab\_graphite}) / A_{interface} \qquad (2)$$

where the slab energies are SBC calculations of either lithium carbonate or graphite (one side hydrogenated) super-cells in the same composition as used in the total interface super-cell. All three super-cells have been structurally relaxed.

The formation energy of Li-related defects ($E_f^d$) is defined as:

$$E_f^d = E_{defected} - E_{pristine} \pm \mu_{Li} \qquad (3)$$

where $E_{defected}$ and $E_{pristine}$ are the total energies of the optimized structures with or without defects respectively. The lithium chemical potential ($\mu_{Li}$) is fixed to be the one of metallic lithium.

### 3. Results and discussions.

While looking for potentially new stable structures, the best choice is often to use exhaustive potential energy surface strategies, like minima-hopping techniques [34-35]. Although their use is theoretically possible, it has never been applied to an interface between materials with different bonding schemes, making the treatment of the interface stoichiometry delicate. Indeed, we are interested in testing various compositions at the interface and thus we propose a different approach based on the comparison of different models. We have used two principal assumptions to construct interface models between graphite and SEI. First, we suppose that the SEI grows from the decomposition product of ethylene carbonate molecules (represented here by lithium carbonate) and forms a crystalline phase [36] on the edge graphite planes. Second, we look for an interface where the preferable Li diffusion channels in bulk $Li_2CO_3$ along the $b$ and $c$-axes [23-25, 37] are aligned with the graphite galleries to minimize the energetic cost of transferring a lithium at the interface.

To assess the proposed models the following parameters are taken into account. First, direct or indirect (*i.e.* through passivation layers) bonding between the graphite edges and the SEI is necessary to ensure their mutual adhesion. Second, the deformations of bound components in the vicinity of the interface should be kept as low as possible to allow lithium to flow easy between the SEI and the graphite. Third, the built system should have low formation energy for thermodynamic reasons. Finally, we expect that the charge distribution due to the ionic nature of $Li_2CO_3$ may induce an electric field at the interface. Hence it is desirable that its direction would assist Li transport toward graphite but not prevent intercalation.

All these criteria were used in the evaluation of five different constructed models (Table

1), which are depicted in details in the Supporting Information (SI). Among them, we found that the interface between graphite and an (*a,b*)-oriented $Li_2CO_3$ slab is the most stable. This interface, shown in Figure 1a, is characterized by tight binding between carbon atoms of the graphite edge and surface oxygen / lithium ions of $Li_2CO_3$. New-formed covalent C-O bonds are an obvious linker at the interface as carbon atoms are present in both compounds and have an equivalent covalent type. At the interface, ionic Li···C contacts [38] compensate the Li···O missing bonds due to cutting of the slab from the bulk lithium carbonate phase. As a result, the formed C-O and C···Li bonds terminate all dangling bonds at the interface and maintain the steric environment peculiar to the crystal organizations of both materials. Namely, mean coordination numbers of the interface carbon atoms and lithium ions perfectly match with those in bulk phases (Table 1) as well as their trigonal planar and tetrahedral environment respectively. Concerning oxygen ions, the mean coordination number 2.3 in the model interface seems to be less satisfied when compared directly with the mean bulk number 3.7 in $Li_2CO_3$. However, oxygen species seem to adapt to various environment as can be seen from crystal organization of $Li_2CO_3$, where two different types of oxygen ions are present either with 3 or 4 neighbors. Hence, at the interface it ensures a more flexible coordination arrangement around the carbonate groups and has less impact to the interface stability than C or Li.

      The good matching between the coordination environment of Li and O ions and the dangling bonds at the graphite edge induces minimal deformations of connected materials at the interface. As a consequence, this model presents the lowest surface formation energy compared to the other models where dangling bonds or under-coordinated ions are present at the interface (see Table 1). As far as adhesion of SEI on top of graphite is concerned (see Eq. 2), this model presents a gain of 116 meV/Å$^2$. The topology of the proposed model corroborates with

crystalline domains of $Li_2CO_3$ registered on graphite anodes as assessed by means of transmission emission spectroscopy [12, 14]. In particular, the observed interplanar distances of 0.28 nm coincide with the size of small channels in crystal $Li_2CO_3$ which are visible from the top view of the model (Figure 1S, SI).

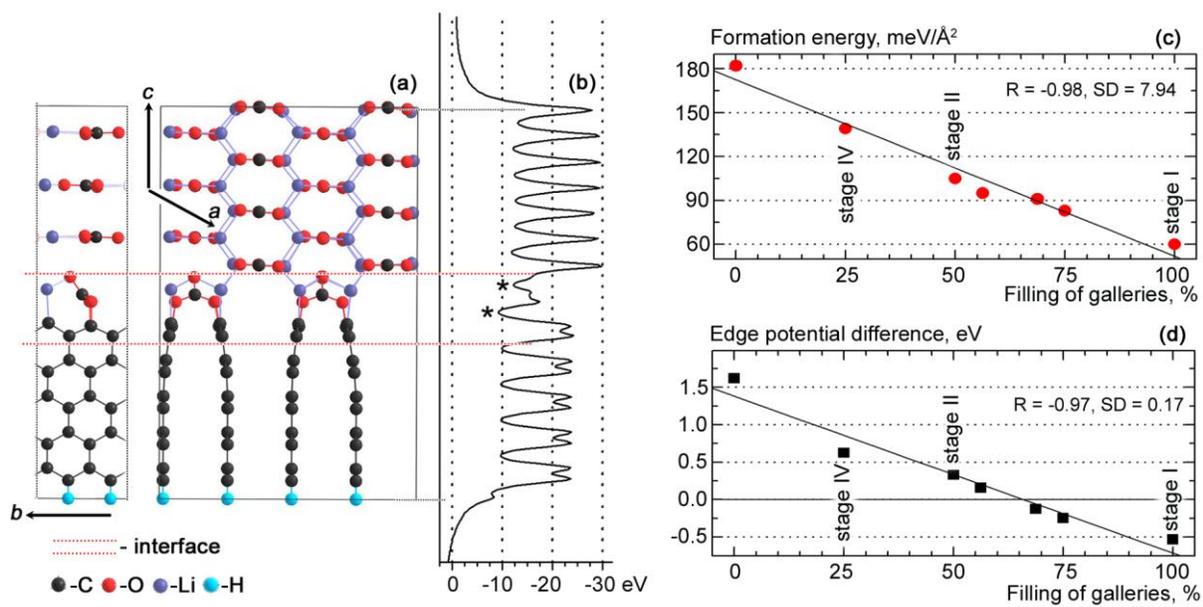

**Figure 1 (Color-on-Line).** (a) Model of the proposed graphite - $Li_2CO_3$ interface. The position of the interface is aligned (red dot lines). Ball and stick representations of the interface in two orthogonal directions after atomic position relaxation in the super-cell. The crystalline character of $Li_2CO_3$ is well preserved while shared carbonate groups at the interface link the two phases together. (b) Projection of the total potential along the $c$-axis. Potential maxima due to the connection of different type compounds are marked by asterisk. (c, d) Formation energy (equation 1) and edge potential difference as functions of gallery filling, R is correlation coefficient, SD is standard deviation.

Table 1. Main characteristics of the proposed models.

|  | Models | | | | | |
|---|---|---|---|---|---|---|
|  | **1**[a] | **2**[b] | **3**[b] | **4**[b] | **5**[b] | **6**[c] |
| Orientation of the $Li_2CO_3$ slab | *a,b* | *a,c* | *a,c* | *a,c* | *a,c* | *a,c* |
| Differences in formation energy, meV/Å$^2$ [d] | 0 | 381 | 274 | 152 | 100 | 203 |
| Edge potential difference, eV [e] | 1.59 | -3.20 | -3.04 | -1.14 | -0.70 | -1.72 |
| Interface potential affinity, eV [f] | 0.41 | 5.22 | 2.30 | 2.65 | 1.93 | 1.45 |
| Mean coordination number, C (3.0) [g] | 3.0 | 3.0 | 2.4 | 1.9 | 2.6 | - |
| Mean coordination number, Li (4.0) [h] | 4.0 | 2.5 | 3.0 | 4.5 | 3.5 | - |
| Mean coordination number, O (3.7) [i] | 2.3 | 1.7 | 2.0 | 2.6 | 3.0 | - |

[a] see Figure 1;

[b] see Figures 2S-5S, SI;

[c] molecular dynamic snapshot, that further relaxed within DFT [37, 39] (see Figures 6S, SI);

[d] see equation 1, model 1 being taken as reference, its value from equation 1 being 182 meV/Å$^2$;

[e] difference between bottom and top edge potential values of the model.

[f] absolute difference between the mean maximum bulk potential value and the mean maximum interface potential value (see Figure 1, 2S-6S, Table 2S);

[g] contacts of the edge graphite carbons with oxygen or lithium ions within the mean bond lengths (1.42 Å, C-O [40] and 2.30 Å, C···Li [38]) are taken into account. The coordination number in bulk graphite is given in parenthesis;

[h] contacts of the edge lithium ions with oxygen or carbon ions within the sum of ionic radii (2.16 Å, O···Li [41]) or the mean bond length (2.30 Å, C···Li [38]) are taken into account. The coordination number for Li in bulk $Li_2CO_3$ [36] is given in parenthesis;

[i] contacts of the edge oxygen ions with lithium or carbon ions within the sum of ionic radii (2.16 Å, O···Li [38]) or the mean bond length (1.42 Å, C-O) are taken into account. The mean coordination number for O in bulk $Li_2CO_3$ [36] is given in parenthesis.

For this model a projection of the total potential along the *c*-axis (Figure 1b) is quite uniform with periodic funnels whose minima correspond to atomic positions (depth is determined by pseudopotentials). For graphite side each funnel with a minimum of about -24 eV splits and origins from superposition of 32 coplanar carbon atoms. For $Li_2CO_3$ side funnels with minima of about -30 eV come from the superposition of 24 coplanar Li, C and O atoms. At the interface the potential is distorted due to out-of-plane deformation of the edge carbonate groups as a result of their partial binding with graphite.

To evaluate the connection between the two materials we define so-called "interface potential affinity with bulk" (Table 1, 2S) that implies to be as small as possible to insure the electronic homogeneity. For this purpose four - five maxima on the potential curve are taken corresponding to the internal parts of the graphite and $Li_2CO_3$ slabs. The precise number of chosen maxima depends on the slab configuration. Next, these maxima are averaged resulting in a mean maximal bulk potential value. A mean maximal interface potential value is calculated as an average of maxima coinciding with contacts of different type groups at the interface (Figure 1b, marked by asterisk). The absolute difference of mean maximal bulk and interface potentials of 0.41 eV demonstrates the high potential kinship between the bulk and the interface for the model 1.

At the same time, the mean potentials for $Li_2CO_3$ and graphite sides (an imagine border between them was taken at the maximal potential point on the curve) are found to be -13.46 eV and -12.91 eV respectively (Table 1S). These values reflect quite similar charge distribution in the two slabs. Nevertheless, small imbalance potentials of -0.81 and 0.78 eV of the top and bottom edges (taken at final points on the potential curve tails in Figure 1b and whose subtraction defines the edge potential difference) induce an electric field opposite to the up-direction of the *c*-axis. The presence of such electric gradient was recently assessed by means of photoelectron spectroscopic examination of the buried interfaces in cycled LIBs [42].

Contrary to the above-described model, all interfaces based on an *(a,c)*-oriented $Li_2CO_3$ slab are much less stable (see models 2-5, Table 1). They are characterized by high formation energies, moderate potential affinities, and strong structural distortions (Figures 2S-5S, SI) as a consequence of unmatched coordination environment and steric hindrance. Moreover, due to non-uniform potentials of the graphite and $Li_2CO_3$ phases and resulting edge potential differences the induced electric fields regardless of their magnitude for all these models are orientated from the bottom to the top edge. That makes the Li diffusion toward graphite galleries difficult. A correlation between the formation energies and the edge potential differences (Table 1) can be directly attributed to the quality of two material connection. Indeed, a weak binding results in low stability with high formation energies and moderate interface potential affinities that must be compensated by higher edge potentials.

When thermal disorder is additionally present in the *(a c)*-oriented slab (model 6 obtained by relaxation of an MD snapshot [39], Figure 6S), the crystalline character of the $Li_2CO_3$ is easily lost and the formation energy is higher than the one of our proposed model by 203 meV/Å$^2$. We assign it to the absence of binding interface contacts, quickly sending disorder in the carbonate phase. Interestingly, the formation energy of the latter model is still lower than those of models 2-3, highlighting that the passivation of graphene edges with carbonate groups disconnected from the $Li_2CO_3$ phase is not an option for interface stabilization. It is also confirmed by evaluations of the interface potential affinity (1.45 eV for the model 6 vs. 5.22 / 2.30 eV for the models 2-3, Table 1). The fact that only carbonate groups shared between graphite and $Li_2CO_3$ are favored from the energetic evaluations supports the view of the mechanical role of the SEI to prevent exfoliation of the sheets [43].

Next, we have evaluated the stability of the proposed model 1 upon lithiation. In order to reduce the complexity, a standard stacking for Li is considered instead of the more stable domain packing [44]. It is found, that intercalation of Li ions into the graphite galleries additionally stabilizes the model interface. Indeed, the formation energy of our model is reduced by 43 and 122 meV/Å$^2$ for the lithiated stages IV and I (Figure 1c, stage numbers correspond to empty graphene planes sandwiched between two consecutive lithium planes [44-45], Figure 7S). This stabilizing effect is connected with the crucial role of lithium in maintaining the $\alpha,\alpha$-stacking of graphite, while pure graphite adopts spontaneously the $\alpha,\beta$-configuration. Moreover, the edge potential difference is diminished for the stages IV and II (Figure 1d) due to the improved charge balance under graphite lithiation. When 70% of galleries are already filled, the edge potential difference vanishes and further intercalation induces an electric field of opposite direction counteracting the Li transport. This is attributed to the evolution of the mean potential of the lithiated graphite side that exceeds the mean potential of the $Li_2CO_3$ side after the stage II (-14.62 eV vs -13.17 eV).

Based on the validated model interface its impact on Li transport inside the passivated anode is further studied. We take into account three of the possible diffusion mechanisms either by a Frenkel pair or by interstitials and vacancies. All of them were shown to be probable mediators for Li diffusion in bulk $Li_2CO_3$ in dependence on the applied voltage [23-25]. It must be noted that here we considered only neutral defects generated by displacement, addition or removing of a Li atom.

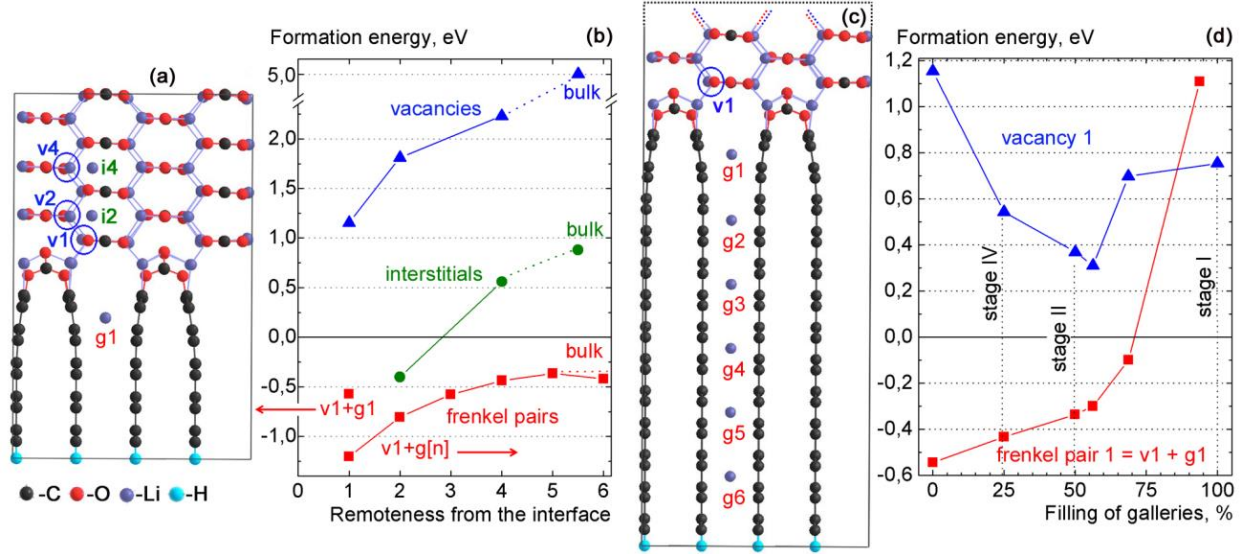

**Figure 2 (Color-on-Line).** Positions of point defects (a, c) and their formation energies (see equation 3) as functions of remoteness from the interface (b) or gallery filling (d). Vacancy (with circles) and interstitial positions in Li$_2$CO$_3$ are depicted by blue and green. The positions of interlaced Li atoms in graphite are indicated by red labels. The same color convention is used to plot the variations of the formation energies in (b) and (d). Note: the graphite slab was extended in (c) to obtain an internal area free from the edge effects [46]. The chosen 2x2x2 dimension of the Li$_2$CO$_3$ slab (a) is already sufficient for surface and internal part differentiation.

With the orientation of the induced potential gradient, the Li intercalation into empty galleries can be initiated by the generation of a Frenkel pair (FP). The preferable positions of the Li vacancy at the Li$_2$CO$_3$ surface (v1) and the Li intercalated between the graphite sheets (g1) are shown in Figure 2a. It is worth noticing that the presence of the interface naturally favors the separation of the pair. Indeed, as can be seen in Figure 2b the depicted FP (v1+g1) with $E_f^d = -0.55$ eV is more stable than the pristine structure or the pure Li$_2$CO$_3$ FP (v1+i2, $E_f^d \sim +0.76$ eV). Such displacement of the interface Li can be carried out by interaction with the

Highest Occupied Orbital as demonstrated in Ref. 47, which is precisely localized at the interface (Figure 1S). A distance of 5.18 Å between two defects agrees well with evaluations for stable FPs in bulk $Li_2CO_3$ [23] highlighting the affinity of the interface with bulk. Presence of several FP defects (Figure 8S, SI) is found to be less profitable in comparison with the pristine structure. Such effect can be attributed to substantial distortion of the interface $Li_2CO_3$ area due to multiple Li displacements into the galleries. Indeed, the latter makes residuary Li ions to shift from their lattice sites and to be shared with the under-coordinated interface oxygen ions. Such loss of structural integrity was observed for the $Li_2CO_3$/Au interface with removal of several Li [48].

The generation of individual point defects such as Li vacancy and interstitial (v1 and i2, Figure 2a, b) cost 1.15 eV and -0.39 eV respectively. These formation energies slowly increase with the remoteness from the interface (v1-v4 and i2-i4, Figure 2a, b) and may reach the bulk values of 4.96 eV and 0.88 eV far away from the graphite [23]. The intercalation energy in graphite is also affected by the interface. It grows if Li is placed deeper into the galleries (v1+g[n], Figure 2b, c). When the Li ion is screened of $Li_2CO_3$ by 8 aromatic cycles binding with graphite saturates and reaches the bulk value -0.37 eV (g4, Figure 2b, c). The slight decrease observed at position g6 is attributed to the vicinity of the bottom edge of graphite that leads to a stronger binding of the Li wave functions with the H-passivated edge [46-47].

The obtained results demonstrate that the defects are accumulated at the interface. We assign this effect to the force induced by the potential gradient. Localization of accessible electrons within the HOMO also facilitates the charge redistribution during defect formation. For example, the interstitial Li atom is usually fully ionized in bulk $Li_2CO_3$ or graphite due to the charge transfer to nearby carbon atoms [23, 49]. Finally, modified coordination environment raises the flexibility of the materials at the interface and makes formation of defects less costly.

Based on these data, we can sketch the following mechanism for the Li intercalation in the presence of a capping carbonate layer on top the graphite anode. First, some lattice Li ion available in the interface carbonate layer diffuses into a gallery driven by the electric potential gradient. The concentration of such intrinsic Frenkel pairs at the interface is evaluated to not exceed seven defects per 1000 Å$^2$. Then, the corresponding vacancy can be occupied by a nearby Li interstitial accumulated at the interface, in the vicinity of which it has a negative formation energy. Such interstitials are coordinated into big channels aligned with the *b*-axis and move through a knock-off mechanism with participation of the lattice Li ions, *i.e.* the principal mechanism at works for interstitial diffusion in bulk Li$_2$CO$_3$ [23-24]. Defect annihilation can be also provided by neighboring lattice Li ions through a direct hopping mechanism [23, 25], but it seems less probable due to the much higher formation energy of vacancies in comparison with the interstitials. A possible decrease at the interface of diffusion barriers and resulting in favored diffusion seems negligible, because a barrier for v1-v2 transition was evaluated to be about 0.3 eV that coincides with the bulk value [25].

As can be seen in Figure 2d, such intercalation mechanism is favorable until galleries are filled up to 70% although the formation energy gradually increases. For almost filled graphite the intercalation by the FP mechanism becomes difficult with a positive formation energy of ~1 eV. Indeed for higher intercalated states the spontaneous Li displacement is not supported by the potential gradient and the lattice ion has to be pushed out by the interstitial Li within the mentioned knock-off mechanism [23-24]. Vacancies are supposed to play a minor role as mediators due to their relatively high formation energy. The latter slightly decreases with the filled galleries (Figure 2d) due to balanced charge distribution and weaker induced potential gradient between the two phases (Figure 1d).

### 4. Conclusions.

Several lithium carbonate - graphite interface models were considered as a prototype of the SEI capping layer on graphite anodes in lithium-ion batteries. Within the Density Functional Theory framework it was demonstrated that only the ($a,b$)-oriented $Li_2CO_3$ slab promotes tight binding with graphite. The found steric organization of the components combines the structural features of both materials and reproduces coordination environment of ions in reference crystal states. As a result, this model is characterized by much lower formation energy and higher potential affinity with bulk than all model interfaces based on the ($a,c$)-oriented $Li_2CO_3$ slab. The charge distribution resulted from localization of ions induces an electric potential gradient at the interface that corroborates with recent experimental data. Lithiation of graphite galleries has an additional stabilizing effect on the model interface.

To clarify the Li transport in the vicinity of the interface we studied the generation of different point defects as possible mediators. It is found that Li diffusion is mainly provided by interstitials wherein the interface traps point defects. The built-in electric field assists the migration of defects through $Li_2CO_3$ to graphite when graphite lithiation ratio is less than 70%. For higher intercalated states a counteracting electric field at the interface is revealed.

Beside the studied inorganic $Li_2CO_3$ component, the obtained results can be generalized to other species of the SEI. We propose the approach applied here as a general tool for such interface research. Indeed, the low surface formation energy, balanced coordination environment and the interface potential affinity are shown to be the key descriptors of the interface stability. Hence they can be recommended as the principal assessments for future interface design.

**Acknowledgements.**

The authors would like to thank Dr. Hakim Iddir (Argonne National Laboratory) for providing the results of MD/DFT simulations of the model 6. The financial support from the European Union (FP7/2007-2013) under grant agreement N 608575 is gratefully acknowledged. This work was granted access to the HPC resources of IDRIS under the allocation 2016-096107 made by GENCI.

**Supporting Information.**

The Supporting Information concerning structural and computational details is available free of charge on the journal's website.

**Technical tools.**

Ball-and-stick pictures and diagrams presented in the paper were created with the V_Sim (http://inac.cea.fr/L_Sim/V_Sim/) and Xmgrace (http://plasmagate.weizmann.ac.il/Grace) software.